\documentclass[%
 reprint,
 amsmath,amssymb,
 aps,
]{revtex4-2}
\usepackage[T1]{fontenc}
\usepackage[utf8]{inputenc}
\usepackage{cmap}
\usepackage{amsmath}
\usepackage{amssymb}
\usepackage{amstext}
\usepackage{amsfonts}
\usepackage{mathtools}
\usepackage{hyperref}
\usepackage{todonotes}
\usepackage{physics}
\usepackage{amsthm}
\usepackage{enumitem}
\usepackage{graphicx}% Include figure files
\usepackage{dcolumn}% Align table columns on decimal point
\usepackage{bm}% bold math
\usepackage{hyperref}% add hypertext capabilities
\usepackage[capitalise]{cleveref}
\hypersetup{
 colorlinks=true,
 linkcolor=blue,
 citecolor=blue,
 filecolor=blue,
 urlcolor=blue,
 anchorcolor=blue,
 bookmarksopen=true,
 bookmarksopenlevel=1,
 pdfpagemode=UseOutlines,
 pdfstartview={XYZ null null 1},
 linktocpage=true,
}

\usepackage{xcolor}

\begin{document}

\title{Snowmass White Paper: \\ 
Quantum Computing Systems and Software for High-energy Physics Research}% Force line breaks with \\

\thanks{This manuscript has been authored by UT-Battelle, LLC under Contract No.~DE-AC05-00OR22725 and Lawrence Berkeley National Laboratory under Contract No.~DE-AC02-05CH11231 with the U.S. Department of Energy. The United States Government retains and the publisher, by accepting the article for publication, acknowledges that the United States Government retains a non-exclusive, paid-up, irrevocable, world-wide license to publish or reproduce the published form of this manuscript, or allow others to do so, for United States Government purposes. The Department of Energy will provide public access to these results of federally sponsored research in accordance with the DOE Public Access Plan ({http://energy.gov/downloads/doe-public-access-plan}).}%

\author{Travis S.~Humble}
\email{humblets@ornl.gov}
\author{Andrea Delgado}%
\author{Raphael Pooser}
\author{Christopher Seck}
\author{Ryan Bennink}
\author{Vicente Leyton-Ortega}
\author{C.-C.~Joseph Wang}
\author{Eugene Dumitrescu}
\author{Titus Morris}
\author{Kathleen Hamilton}
%%\author{Phil Lotshaw}
\author{Dmitry Lyakh}
\author{Prasanna Date}
\author{Yan Wang}
\author{Nicholas A.~Peters}
\author{Katherine J.~Evans}
\author{Marcel Demarteau}
\affiliation{%
 Oak Ridge National Laboratory, Oak Ridge, Tennessee, USA}%
\author{Alex McCaskey}
\affiliation{NVIDIA, Santa Clara CA, USA}
 \author{Thien Nguyen}
 \affiliation{Quantum Brilliance, The Australian National University, Canberra, Australia}
 \author{Susan Clark}
 \author{Melissa Reville}
 \affiliation{Sandia National Laboratories, Albuquerque, New Mexico, USA}
 \author{Alberto Di Meglio}
 \author{Michele Grossi}
 \author{Sofia Vallecorsa}
 \affiliation{CERN - European Organization for Nuclear Research, Geneva, Switzerland}
 \author{Kerstin Borras}
 \altaffiliation{Also at RWTH Aachen University, Aachen, Germany}
 \author{Karl Jansen}
 \author{Dirk Kr\"ucker,}
 \affiliation{Deutsches Elektronen-Synchrotron - DESY, Hamburg and Zeuthen, Germany}

\date{\today}% It is always \today, today,
       % but any date may be explicitly specified

\begin{abstract}
Quantum computing offers a new paradigm for advancing high-energy physics research by enabling novel methods for representing and reasoning about fundamental quantum mechanical phenomena. Realizing these ideals will require the development of novel computational tools for modeling and simulation, detection and classification, data analysis, and forecasting of high-energy physics (HEP) experiments. While the emerging hardware, software, and applications of quantum computing are exciting opportunities, significant gaps remain in integrating such techniques into the HEP community research programs. Here we identify both the challenges and opportunities for developing quantum computing systems and software to advance HEP discovery science. We describe opportunities for the focused development of algorithms, applications, software, hardware, and infrastructure to support both practical and theoretical applications of quantum computing to HEP problems within the next 10 years.
\end{abstract}

%\keywords{Suggested keywords}%Use showkeys class option if keyword
               %display desired
\maketitle

%%\tableofcontents

\section{\label{sec:motivation} Motivation}

The aim of high-energy physics (HEP) is to understand matter at the most fundamental level. The current understanding of the building blocks of the universe and their interactions is embodied in the so-called standard model (SM) of particle physics. In this model, fundamental particles are quantum mechanical entities with fixed quantum numbers such as mass, spin, electric, and color charge, and are regarded as excitations of relativistic quantum fields. While the field of HEP has enjoyed many decades of progress, a long-standing challenge for both theorists and experimentalists in HEP is that the properties and behaviors of perturbative quantum fields are mathematically complex, present infrared and ultraviolet divergence challenges, and expansions converge slowly when the coupling constant is large. Furthermore, the time evolution of final state particles and their hadronization is particularly complex and are difficult to accurately simulate on digital computers, even on high performance computer (HPC) systems. 

Meanwhile, a new approach to computing that explicitly leverages the complexities of quantum mechanics for computational purposes has been in development. In the last few years, prototype quantum computers capable of performing small computations using a few dozen quantum bits (``qubits'') have become widely accessible, prompting a burgeoning field of research in applications of near-term and future quantum computers.
Quantum computing (QC) approaches have been developed and demonstrated for applications in many disciplines including chemistry, materials science, nuclear physics, and HEP. However, the field of QC is far from mature and the potential of QC for HEP has only begun to be explored.

In this paper, we highlight some of the significant opportunities and challenges for the development of quantum computing systems and associated software to deliver unprecedented capabilities for HEP discovery science.
In Sec.~\ref{sec:background}, we survey current approaches to applying QC to HEP.
In Sec.~\ref{sec:challenges}, we describe the leading technical challenges and gaps facing the adoption of quantum computing for HEP research problems, while in Sec.~\ref{sec:prior}, we respond with long-term research priorities that will address these gaps in the coming decade.

\section{Background}
\label{sec:background}

Quantum computing may be viewed as part of the larger subject of quantum information science and technology (QIST), which concerns the use of quantum systems to store, transmit, and process information \cite{humble2019quantum}.
QIST involves the precise isolation and control of well-defined quantum particles and fields.
These quantum physical systems can be used to simulate naturally occurring quantum phenomena (such as quantum chromodynamics (QCD)) or to solve computational tasks unrelated to physical phenomena (such as clustering or regression on a dataset).
As will be discussed later in this section, key concepts in quantum computing such as superposition, entanglement, and computational complexity have also provided new, information-oriented ways to address fundamental questions in HEP.

For the most part, efforts to apply QC in HEP to date have consisted of adapting established general-purpose quantum algorithms and tools to HEP-specific tasks. While most demonstrations have involved only toy problems, there have been a few demonstrations in which the quantum computations were competitive with more traditional methods. Some examples include the simulation of a full SU(2) gauge theory on a noisy, intermediate-scale quantum (NISQ) device \cite{Atas2021SU2}, the training of large-scale quantum machine learning (QML) models for supernova classification \cite{Peters2021}, and the generation of synthetic detector data \cite{delgadoHamilton2022unsupervised}. 

\par 
In the United States, much of the work to date exploring the potential of quantum information science for HEP has been through the Department of Energy (DOE) QuantISED program. QuantISED was launched in 2018 as part of the DOE Office of Science (SC) QIS initiative following a series of community roundtables, pilot studies, and 
\cite{DOESensorsReport,GrandChallengesReport}. The program has relied upon interdisciplinary collaboration between HEP and QIS researchers and partnerships between DOE laboratories, university consortia, and hardware vendors such as IBM, Rigetti, and Google. Major topic areas addressed by QuantISED include:

\begin{itemize}
  \item \textbf{Cosmos and qubits:} theoretical connections between cosmic physics and qubit systems that can be studied in laboratory settings. In particular, the study of quantum gravity in the context of quantum information has yielded new insights. For example, principles of computational complexity and quantum error correction have helped scientists investigate the interiors of black holes \cite{Kim2020} and whether gravity emerges from entanglement \cite{Periwal2021}.
  
  \item \textbf{Foundational theory:} formulations of gauge theories amenable to simulation on quantum computers \cite{PhysRevD.104.094519}.
  
  \item \textbf{Quantum simulation:} algorithms for simulating quantum mechanical systems including scalar quantum field theories \cite{yeter2019scalar}, nuclear physics \cite{yeter2020practical}, and other many-body systems \cite{yeter2021scattering}.
  
  \item \textbf{Quantum computing:} quantum-enhanced machine learning and data analysis for HEP experiments. Developments in this area include: quantum machine learning algorithms \cite{Peters2021, 2021Wu, 2021WuSunGuan} and optimization algorithms for the analysis of collider data and quantum simulation of field theories \cite{QASimNachman2021}.
  
  \item \textbf{Quantum sensors:} sensors leveraging quantum information concepts and technologies to yield new detection capabilities. One advancement in this area is quantum-enhanced searches for dark matter \cite{QISforAxions, PhysRevLett.126.141302}.
  
  \item \textbf{Quantum technology:} advances in QIST and its integration with HEP technology. Efforts in this topic include quantum internet \cite{PRXQuantum.1.020317}, noise mitigation techniques for near-term quantum devices \cite{UnfoldingNachman2020}, analysis of the technological requirements for HEP programs \cite{PhysRevD.103.054507}, and others.
\end{itemize}

While there is not space to describe all of these efforts in detail, recent work relating black hole thermodynamics and quantum information theory exemplifies the kinds of fruitful connections being developed between QIS and HEP. 
A long-standing problem in particle physics is the so-called ``black-hole information paradox'', which concerns the apparent loss of information that occurs when a particle disappears inside a black hole. The particle's information chaotically mixes with all the other matter and information inside the black hole, generating quantum entanglement between distant regions and seemingly making it impossible to retrieve. Recent landmark work \cite{Almheiri2019,Penington2020} suggests that the information passing the event horizon of a black hole is not lost forever but is eventually released.

In the field of quantum information, this process of apparent information loss through widespread entanglement is known as scrambling. A quantum processor can directly measure the scrambling-induced spread of initially localized information via the decay of out-of-time-ordered correlation (OTOC) functions. In \cite{blok_quantum_2021}, the authors realize scrambling behavior in a multi-qutrit system. In \cite{Landsman2019}, a seven-qubit circuit executed on an ion-trap quantum computer enabled the authors to bound the scrambling-induced decay of the OTOC measurement experimentally.
These two experiments pave the way for laboratory investigations of the black hole information paradox, something that would hardly be imaginable without QC. 

\section{\label{sec:challenges} Challenges}
The emerging field of QC faces many technical challenges related to the development of algorithms, software, hardware, and infrastructure to support HEP research. In this section, we identify how these existing research challenges influence on-going efforts to solve key problems in HEP research using QC. In Sec.~\ref{sec:ana}, we survey algorithms and applications which are known to apply to HEP challenge problems, and we identify several key gaps in extending these methods beyond proof of principle efforts. In Sec.~\ref{sec:sah}, we review how existing computational tools, including quantum programming methods and numerical simulators, have influenced the accessibility and adoption of QC for the HEP community. In Sec.~\ref{sec:tfh}, we briefly summarize early efforts in developing testbeds for experimental QC with a focus on HEP application areas. Then, in Sec.~\ref{sec:infra}, we discuss some of the broader infrastructure challenges that face adoption of QC by the HEP community. 
\subsection{Algorithms and Applications }
\label{sec:ana}
The Hamiltonian formulation of lattice quantum field theories allows us to use the tools from quantum computation to solve problems in strongly correlated quantum systems by understanding the dynamics in terms of quantum circuits or representing the action of a measurement as a projection in a Hilbert space \cite{Banuls2020,Buser2021}. Different approaches in the quantum computation domain for quantum field theories cover subatomic many-body physics \cite{jordan2012quantum,lu2018simulations}, real-time model dynamics of lattice gauge theories \cite{martinez2016real,muschik2017u}, self-verifying variational quantum simulation of the lattice Schwinger model \cite{kokail2018self}, Non-abelian SU(2) lattice gauge theories on superconducting devices \cite{mezzacapo2015non}, optical abelian and simulations of non-abelian gauge theories on ultra-cold atoms platforms \cite{tagliacozzo2013optical,tagliacozzo2013simulation}, and simulation of Z(2) lattice gauge theories with dynamical fermionic matter \cite{zohar2017digital,zohar2017digitalA}. 
\par 
In most of these applications, the limited number of qubits and gate fidelities available in QC hardware implementations represent bottlenecks for a proper comparison with classical algorithms and, therefore, any assessment of quantum advantage. On the other hand, the application of QCD with the lattice gauge model \cite{kan2021lattice, stetina2022} represents an exciting challenge on the real-device implementation that requires a novel design following quantum device limitations, state-of-the-art classical simulators handling a decent number of qubits to benchmark the quantum computation algorithms with existing classical approaches, and the exploration of gravity/QFT duality \cite{Buser2021}.
\par
From an algorithmic point of view, several challenges address the exploration of variational quantum algorithms in a higher dimension, design of quantum algorithms beyond the existing hybrid classical-quantum format with the combination of different strategies to solve problems in HEP, exploitation of quantum singular-value-decomposition (SVD) methods for unfolding problems in HEP, and on-hardware implementation of quantum algorithms to reconstruct Feynman path integrals. 
\subsubsection{Digital and Analog Simulations}
Interestingly, digital, analog, and mixed digital-analog models of quantum computation have been considered within the context of lattice gauge theories \cite{RevModPhys.86.153,PRXQuantum.2.017003}. In the digital realm the problem of a \textit{gauge-invariant} problem formulation is already a non-trivial matter with quantum-error-correction-like protocols \cite{stryker2019} and manifestly gauge-invariant formulations \cite{Davoudi2021,Ciavarella2021} having been considered. After overcoming the first hurdle of problem formulation the ever-growing set of quantum algorithms, including quantum phase estimation and linear combinations of unitaries or block encoding, can be used to compute the various quantities of interest appearing in the target theory \cite{Roggero2019,Roggero2020, Ciavarella2021}. These algorithms seem suitable for fault tolerant computation rather than near term quantum computational models.

Analog approaches have long been taken with synthetic gauge fields in cold atom simulators being a prime example of analog simulation \cite{Galitski2019}. Analog realizations of gauge invariant encodings, such loop-string-hadron encoding in cold atom systems \cite{Dasgupta2022}, have also appeared. Lastly, with some modest abstraction one popular near term approach has been to leverage the natural device physics in order to augment the conventional digital gate-sets with analog processes \cite{PhysRevA.104.042602}. For example, select bosonic vibrational modes can be used to simulate $U(1)$ gauge variables in a resource efficient manner \cite{tong2021provably}, e.g., using trapped ion systems \cite{nguyen2021digital}.
\subsubsection{Variational Methods}
Generative modeling learns how to generate a distribution by analyzing a known data. The model itself may then be used to draw additional samples representative of the distribution. With a quantum processor, preparing and sampling from arbitrary distributions can be directly implemented, and there is no need to compute partition functions which may be intractable. Additionally, using quantum systems has an attractive scaling characteristic: a model built on $n$ qubits can potentially prepare and sample from a distribution over $2^n$ states, where each state is a length-$n$ binary bitstring. 

In quantum machine learning there is a broad field of research which includes using classical machine learning algorithms for studying and simulating quantum systems \cite{gao2017efficient,torlai2018neural,torlai2020machine} and using quantum algorithms to implement analogues of classical machine learning algorithms. Many machine learning applications in HEP leverage generative modeling \cite{salamani2018deep,di2019dijetgan,alanazi2020simulation,hariri2021graph}, and the leading model is the generative adversarial network (GAN). Quantum GAN models \cite{dallaire2018quantum,zoufal2019quantum,chakrabarti2019quantum} have been developed that utilize a trainable quantum circuit model for the generator network, yet adversarial training can still encounter mode collapse as with classical models. However, there are other generative models that can be constructed with quantum circuits including quantum Boltzmann machines \cite{amin2018quantum,zoufal2021variational} and quantum circuit Born machines (and related Ising Born Machines) \cite{benedetti2019generative,coyle2020born}.

Near-term demonstrations of QML can reach comparable performance with, but have not yet demonstrated an advantage over, classical machine learning models. Classical machine learning and deep learning can utilize large, distributed computing clusters to implement models with millions of parameters. Coupled with parallelization, these training workflows dwarf the current state of the art in quantum machine learning that uses variational, hybrid algorithms. However, as the size and capabilities of available quantum devices improve, larger quantum models can be built and executed. While near-term demonstrations and applications may not outperform classical ML models, their development is essential for understanding how quantum models learn.

\subsubsection{Noise Mitigation}
The design space associated with quantum generative models, and parameterized quantum models in general, is large. Best practices for building parameterized quantum circuit models, in order to effectively prepare and sample from these distributions, remains an open question \cite{nakaji2021expressibility}. Many heuristic approaches abound including: using sparse parameterizations (single rotation gates instead of arbitrary three angle decompositions) reduces training overhead, at the expense of trainability; and using fewer entangling and two-qubit operations can reduce susceptibility to hardware noise and improve transpilation for sparse hardware connectivity graphs, but at the expense of the correlations that can be modeled. 

Mitigating hardware noise requires efficient noise characterization \cite{PhysRevA.103.042603,dahlhauser2022benchmarking}. Many methods of error mitigation that rely on regression modeling (e.g. \cite{lowe2021unified}) are focused on mitigating the effect of noise on expectation values. For generative modeling and algorithms that use the distribution over all basis states, matrix-based error mitigation methods have been incorporated into variational training workflows \cite{hamilton2020error}, and scalable error mitigation methods are an active area of research \cite{hamilton2020scalable,nation2021scalable}. These methods are needed to assess the stability of the quantum devices as well as reproducibility of the applications \cite{dasgupta2022characterizing}.

\subsubsection{Quantum Machine Learning}
Modern machine learning (ML) techniques, including deep learning, are rapidly being applied, adapted, and developed for HEP applications. ML is currently used across the different areas of particle physics from collider physics \cite{Stakia2021, Mikuni2021, Nachman2021} to the study of the cosmos \cite{Ostdiek2022, List2021} and quantum gravity. ML is also present throughout the various stages of HEP studies from the simulation of hypothesized particles \cite{brehmer2020simulationbased}, in the parameterization of cross-sections calculated for sensitivity analyses \cite{Carrazza2021}, to the analysis of experimental data \cite{Stakia2021,Mikuni2021,Nachman2021}. Most of these applications have replaced non-ML methods and techniques and can process experimental data very efficiently. These methods and techniques have even been deployed in FPGAs \cite{Iiyama2021} and integrated into the data acquisition systems for fast selection of information. 

More recently, applications at the intersection of quantum computing and machine learning have been explored to analyze experimental data. The hope is to adapt or develop algorithms that can efficiently process HEP data. One example includes algorithms to construct physics objects amenable to analysis from the signals generated in a particle detector--i.e., the clustering of detector hits into so-called tracks for reconstructing a particle's trajectory \cite{magano2021quantum, Zlokapa2021, Bapst2019, Shapoval2019, das2020track, Tysz2021, 2020Cenk, Quiroz2021} or tracks and calorimeter energy depositions into jets \cite{Wei2020, pires2020adiabatic, pires2021digital}. Furthermore, quantum-assisted algorithms have been explored in unsupervised learning settings to classify jets according to their origin (b-tagging) \cite{gianelle2022quantum}, generative tasks \cite{bravoprieto2021stylebased, Chang2021, perez2021determining}, and the selection of events or interactions along with background suppression \cite{Mott2017, qamlz, kim2021leveraging, Peters2021, caldeira2020restricted, Belis2021, Terashi2021, AlexiadesArmenakas2021, Bargassa2021, matchev2020quantum, Blance2021, Blance2021a, 2021Wu, chen2020quantum, Heredge2021}. In particular, generative models have been explored extensively as an alternative for the simulation of particle interactions and the detector's response to such interactions \cite{Chang2021, delgadoHamilton2022unsupervised}.

Leveraging quantum computers for machine learning in HEP has several advantages. In general, quantum computers can operate more efficiently in high-dimensional tensor product spaces because of quantum superposition, which is vital for analyzing large-scale and complex HEP data patterns. Moreover, they can leverage quantum tunneling to train machine learning models more efficiently \cite{date2021qubo,date2021adiabatic,arthur2021balanced,date2019classical}. In some cases, quantum-assisted techniques have been shown to train on less data than their classical counterparts. This might be particularly useful in the context of data augmentation in simulation tasks, where using a generative model can reduce the resources allocated for traditional ML methods.

Nonetheless, even though the applications of QML to HEP data are many, there are several challenges associated with the trainability and deployment of these models on NISQ devices. For example, the size and complexity of HEP datasets requires an efficient encoding into quantum states, resulting in a large overhead in pre-processing. Many training workflows serially encode training data and paralellization requiring either multiple devices or large quantum devices that can run multiple circuits with no cross-talk or interference between circuits. Furthermore, there is little understanding on how to tailor ansatz/circuit design for specific applications and a need for error correction/mitigation techniques that can be scalably constructed and incorporated into variational training workflows. 
\subsection{Software and Compilers}
\label{sec:sah}

\subsubsection{Compiling}

Improvements in quantum computers, and their subsequent enabling of increasingly nontrivial simulations of fermionic systems, have pushed the development of novel encoding of fermionic degrees of freedom. The commonly used mappings from these degrees of freedom to that of spin (i.e., the Jordan-Wigner, Bravyi-Kitaev, or parity mappings) in general lead to largely non-local spin Hamiltonians and unnecessarily large qubit  numbers per fermionic mode mapped. These properties drastically compound circuit and measurement complexity of simulations. Towards countering these expenses, several works have shown both general and system specific mappings that outperform Jordan-Wigner and others in terms of locality of spin Hamiltonians as well as the number of fermionic modes per qubit \cite{brayvi2002,derby2021}. 

Although these economic mappings are useful, their utility in the fault-tolerant era may be overshadowed by instead using \textit{logical} fermionic mappings \cite{setia2019,landahl2021,li2018,jiang2019}. These avoid the overhead associated with fermionic-to-qubit mapping followed by the cost of error corrected codes by directly mapping the underlying fermionic symmetries to error correcting code spaces. While the benefit is obvious in terms of qubit and gate overhead, such embeddings require couching fundamental fermionic operations in terms of fault tolerant operations, which will likely encourage simple, more physically motivated implementations of algorithms and compiling of code. 

Explorations of such mappings, and their relative quantum resource efficiency for gauge-invariant HEP problems of interest remain an open, but important, first step. There have been several successful HEP problem implementations, generally focusing on how to truncate gauge degrees of freedom while simultaneously satisfying gauge constraints efficiently \cite{klco2021,Davoudi2021}. This has paved the way, similarly to fermionic mappings, for methods of how to embed HEP simulations natively in error detection and correction schemes \cite{rajput2021,stryker2019}. These methods have so far proven to be very specific to the target models and its symmetries. Thus the endeavor to find general, optimal mappings will be essential in the coming years for scaling up near term HEP simulations into the fault tolerant era.

%% Raphael, Chris
\subsection{Experimental Testbeds}
\label{sec:tfh}
The current state of the art for testbed devices consists of quantum processors that contain dozens of qubits. These machines are capable of circuit depths in the 100s of gates and up to 1000s of gates in the most advanced systems. Alternately, algorithms for simulating HEP problems of interest have been implemented on a comparably smaller number of qubits and circuit depths. 

\par
Given the current level of noise and imperfection in practical realizations of quantum computing devices, efficient numerical simulators capable of running on large-scale heterogeneous (HPC) platforms will continue to play an important role in characterization, verification, and validation of quantum devices, as well as analysis of the quantum algorithms intended for those devices. Probing the potential for quantum advantage will require extensive research and development in classical algorithms for quantum computing simulations in order to make a true boundary where quantum wins over classical. 
\par 
An ability to  utilize existing and future classical hardware efficiently is another dimension of optimization on the classical computing side. Furthermore, the HEP-related quantum simulators will require generalization to the qudit-based representation of quantum devices and algorithms. In order to better understand and cope with the noise and decoherence, analog-level simulations of open quantum device dynamics will become vital for their performance optimization. These simulations will require significant HPC resources coupled with highly optimized computer codes. Yet another challenge is related to loading classical (or quantum) data to quantum devices--a problem that can also benefit from advanced quantum-inspired classical algorithms coupled to large-scale HPC resources. 

\subsection{Instrument and Data Networks}
\label{sec:infra}

%% Andrea
Historically, HEP computing has been performed on sizeable, purpose-built computing systems. These began as single-site computing facilities but have evolved into the distributed computing grids that we use today. Therefore, integration with current heterogeneous computing resources is needed to accelerate the adoption of quantum computing technologies within the HEP community. For example the Fermilab HEP cloud is a portal to diverse computing resources on local clusters, campus farms, grid resources, commercial clouds, HPC centers, and quantum computing resources.

\section{\label{sec:prior} Priorities}
\subsection{Algorithms and Applications}
\label{sec:ana_prior}
%%%
With ever increasing numbers of qubits into the hundreds or thousands, while remaining in the NISQ era, quantum algorithm research for HEP applications to improve/overtake classical methods is an important research focus for the next ten years. Quantum computing for HEP provides a great challenge and important intellectual stimulus for overcoming practical quantum information challenges with big data. 
In addition, quantum computing for simulating lattice gauge theories provides both motivation and a framework for interdisciplinary research towards developing special-purpose digital and analog quantum simulators, and ultimately scalable universal quantum computers.

Another potential area that may yield quantum advantage involves processing the large data sets resulting from  detected events.
For instance, simulations and on-hardware implementations for the unfolding problem can exploit both quantum generative modeling as well as singular value decomposition. 
Recent advances also enabled light-front encoding of relativistic fields and subsequent simulation with variational methods~\cite{kreshchuk_quantum_2020,kreshchuk_light-front_2021}. Encoding techniques allow for large simulations to be mapped onto relatively few qubits via the Lie algebraic method to reduce simulation complexity via gauge invariant and sub-sector encodings pertaining to a symmetry~\cite{klco_quantum-classical_2018}, while computing the quantum gradient on the Reimannian manifold allows for increased circuit depths when complexity reductions have been maximized. Approaches beyond the current hybrid quantum-classical paradigm, which combine efficient quantum algorithms with classical distributed algorithms, may also benefit HEP applications.

% %%%
\subsubsection{Variational Algorithms}
%%%
One advantage to using quantum models for statistical learning tasks is that with $n$ qubits, one can build a parameterized model that can efficiently prepare arbitrary distributions over $2^n$ states. The development of QML models for HEP applications can provide insight into how data can be efficiently transformed in high-dimensional probability spaces. The challenges for these approaches to quantum computing can be organized into two areas: designing circuit ansatz models that can efficiently transform data into high-dimensional quantum Hilbert spaces, and building scalable training methods that can optimize these models in noisy landscapes. 

Many state of the art quantum algorithms in the HEP application space that can be deployed on near-term hardware are variational algorithms: these are hybrid workflows that utilize quantum and classical processors to train parameterized quantum models. Training a variational model is a non-convex optimization problem, and the training task can be NP-hard \cite{bittel2021training}. In classical machine learning, gradient-based training methods can be scaled to large-scale applications in non-convex and convex optimization, and efficient training of deep learning models has been facilitated through the use of backpropagation. Analogous methods for variational workflows have been developed \cite{verdon2018universal,beer2020training}, but have not been demonstrated yet on hardware. 

Currently, gradient-based and gradient-free optimization are the leading approaches to training. Gradient-based training is predicted to improve convergence \cite{napp2021gradient}, yet developing gradient-based optimization routines that can effectively incorporate queue-based access to quantum hardware should be a priority for near-term research. As the number of parameters in a circuit increases, and the number of qubits in the register increases, effective scheduling of the experiments that are needed in order to execute one step of an optimization method is needed. Hardware access which allows for job batching (where a set of experiments sent by a single user are executed and results are returned as a single group) can facilitate efficient training. With batched jobs, it may be possible to execute all circuits needed for one optimization step, either evaluation of the loss function gradient (e.g. for gradient descent \cite{liu2018differentiable,mitarai2018quantum,hamilton2019generative}), executing a direct search (e.g. for particle swarm \cite{zhu2019training} or coordinate descent \cite{delgadoHamilton2022unsupervised}), or for gradient approximation (e.g. using finite differences or SPSA \cite{spall1992multivariate}). Additionally, a larger number of shots are needed to efficiently sample from $n$-qubit states. 

\subsubsection{Noisy Algorithms}
%%%
The challenges of effective state preparation, circuit training in noisy landscapes are not mutually exclusive. One observed effect of hardware noise is the loss landscape flattening and the existence of ``barren plateaus'' \cite{mcclean2018barren,wang2021noise}. In a flattened landscape, the efficacy of gradient-based training will be reduced. However, the impact of barren plateaus can be mitigated with circuit ansatz and cost function design \cite{cerezo2021cost,pesah2021absence,uvarov2021barren}.

There are a diverse set of error mitigation methods developed for short depth circuits \cite{temme2017error}. But highly expressible variational models may not be short depth and variational training is not guaranteed to prepare well-verified states. An alternate approach to error mitigation is to use matrix-based methods where sparse linear filters are used to reduce spurious counts in prepared states. Matrix-based error mitigation methods can be incorporated into variational training methods as a data post-processing state, as the added cost of preparing the linear filter.  

\subsubsection{Machine Learning Algorithms}

The known challenges for quantum machine learning (QML) in the HEP application space can be organized into three areas: embedding data into quantum Hilbert spaces, training quantum models in noisy landscapes, and working with quantum data. Currently, the capabilities in classical distributed computing make it unlikely that a quantum advantage will be observed with classical data, or classical representations of quantum data, using QML models deployed on near-term hardware with < 100 qubits. However, the development of QML models for HEP applications can provide insight into how data can be efficiently transformed in high-dimensional probability spaces. 

These three challenges are not mutually exclusive. For example the challenges of training quantum models are encountered irrespective of the data source (classical data, classical representation of quantum data, or quantum data). Examples of how challenges of embedding data can manifest: finding low-dimensional representations of classical data into qubit registers, or high-fidelity transmission of quantum data from sensors to qubits used for QML. 
In addition to these, it is important in QML to explore purely quantum approaches for both training and inferencing, i.e. approaches that are not hybrid in nature (such as variational approaches).
Working with the NISQ-era machines in the near-term, it is also important to explore quantum error-correction and error-mitigation subroutines specific to QML circuits used in HEP.
From a logistical point of view, accessing quantum computers requires better scheduling policies and remote access protocols.
Lastly, today's quantum computers are noisy and error-prone---this severely limits our ability to work with large-scale data reliably.
In this regard, it is important to build fault tolerant quantum computers that are also scalable.

\subsection{Software and Compilers}
  
As discussed in Sec.~\ref{sec:ana} and Sec.~\ref{sec:ana_prior}, hybrid quantum algorithms, such as those based on variational principles, have emerged as a promising candidate for quantum computational advantage in HEP areas, especially in the Noisy Intermediate-Scale Quantum (NISQ) era~\cite{preskill_nisq}. The iterative nature of these algorithms necessitates a tight integration of quantum hardware with classical computing resources. For instance, the execution time of quantum circuits that involves a variational parameter loop has been shown to be dominated (more than 90\%~\cite{ibm_clops}) by classical computing procedures for compilation, control, and data transfer. As quantum computing capabilities mature, software infrastructures for compilation and control will necessarily need to improve in order to utilize the full computing power of heterogeneous quantum-classical resources for handling practical applications.

We envision the need for system-level and hardware-agnostic compiler toolchains akin to those we have today for classical computing - extensible, modular systems with unified intermediate representations that enable a wide array of optimization and code-generation techniques for multiple source languages or target architectures \cite{McCaskeyICRC2018}. The most important aspect of any compilation toolchain design is the Intermediate Representation (IR), which is how code is represented in the compiler. A standardized, robust, and forward-looking quantum-classical IR is, therefore, an essential step towards ensuring full interoperability within the quantum software ecosystem, promoting contributions from broad communities, and enabling optimization and transformation of quantum programs that we discuss in detail in the later point.

For instance, multiple community-led efforts~\cite{qedc_web, qir_alliance_web} in the field have coalesced around the idea of leveraging existing classical compiler infrastructures, like the LLVM, for quantum computing. The approach ensures the low-level coupling of quantum computation with classical computing resources and also supports integration with classical languages and tools readily available in those compiler toolchains. For example, the Quantum Intermediate Representation (QIR) Alliance~\cite{qir_alliance_web} is an open-source community within the Linux Foundation focusing on enabling an LLVM-based IR specification for quantum computation (QIR), as well as developing tooling around this unified representation. Oak Ridge National Laboratory is a founding member of this organization along with several industry players, including Microsoft, Quantinuum, Rigetti, and Quantum Circuits Inc. Importantly, by leveraging LLVM for quantum compilations we could also benefit from its powerful tools, especially the Multi-Level Intermediate Representation (MLIR)~\cite{lattner_mlir} library, which is designed for heterogeneous hardware and domain-specific languages analogous to that of quantum-classical computing paradigm. In this regard, we have successfully demonstrated a prototype compiler toolchain, QCOR~\cite{qcor}, capable of compiling the novel OpenQASM version 3~\cite{openqasm3} quantum programming language to LLVM IR adhering to the QIR specification adopting the multi-stage, progressive lowering approach of MLIR. This compilation technology is specifically relevant to many HEP application use cases whereby the algorithmic procedure is best captured by high-level domain-specific descriptions, such as analog and digital quantum simulation. 

An essential feature of compiler toolchains is the ability to apply transformations on the IR to improve execution quality, e.g., code's runtime or the computing resources required, while preserving the semantics of the input program. In compiler technology, such transformations are often called \emph{passes} since they can be assembled into a pipeline of passes each of which processes the IR and applies its transformation rule.

In the NISQ era of quantum computing, compiler passes that implement circuit simplification and efficient hardware topology mapping are indispensable to any quantum software stack. A broad variety of techniques have been developed quantum circuit simplifications at the gate level, such as those based on ZX-calculus~\cite{van_zx, zx_opt}, template-based~\cite{template_opt} or peep-hole~\cite{nam2018automated} optimizations. Similarly, various circuit-to-hardware placement methods have been developed to minimize the number of qubit swapping operations~\cite{staq, sabre} required or to maximize circuit execution fidelity by mapping the circuit to best-performing qubits~\cite{triq}. 

For dynamical Hamiltonian simulation algorithms that are particularly relevant to HEP applications, techniques such as advanced term splitting~\cite{Hatano2005} and ordering~\cite{childs2019faster, tranter2019ordering} or algebraic circuit compression~\cite{camps2021algebraic} can also be incorporated as optimization passes targeting the high-level description of the quantum algorithms where algorithmic semantics of the program are expressed. In this respect, a quantum compiler infrastructure that supports a hierarchy of abstractions, such as the MLIR, is highly desirable since it can retain high-level algorithmic information in the IR which compiler optimizers can reason \cite{nguyen2021quantum}.  

As quantum computers evolve into fault-tolerant machines, the associated software infrastructure will also need to be upgraded in order to handle quantum error correction protocols \cite{humble2021quantum}. Specifically, we envision that quantum programs would always be expressed in terms of logical operations whose corresponding physical qubit level instructions incorporating a target-dependent quantum error correction code are compiled by a compiler software infrastructure \cite{britt2017high}. In this regard, quantum compilers leveraging classical computing technology, such as LLVM, would be best suited to performing this logical-physical transformation thanks to their ability to incorporate a variety of classical computing resources required by the error decoding and correction protocol, e.g., minimum weight perfect matching~\cite{dennis2002topological} or maximum likelihood decoding~\cite{bravyi_mld} for surface code~\cite{fowler_surface_code}. Furthermore, as quantum programming languages move towards this fault-tolerant paradigm whereby qubit measurement-based control flow is intertwined with that of conventional computing for classical data, we can leverage the vast amount of IR optimization capability from the LLVM infrastructure to handle common classical optimization passes like function inlining or loop unrolling. While these optimization passes are intrinsically classical, they help unlock many quantum-related simplification patterns which would otherwise be obscure to the quantum optimizer due to the complex nature of the IR tree with control flows.
% \end{itemize}
\subsubsection{Encodings}
%%%
The HEP-specific encodings discussed in section II are an important first step to solving gauge-invariant HEP problems of interest in a resource-efficient way.
Explorations of such mappings, and their relative quantum resource efficiency for gauge-invariant HEP problems of interest remain an open, but important first step. There have been several successful HEP problem implementations, generally focusing on how to truncate gauge degrees of freedom while simultaneously satisfying gauge constraints efficiently \cite{klco2021,Davoudi2021}. This has paved the way, similarly to fermionic mappings, for methods of how to actually embed HEP simulations natively in error detection and correction schemes \cite{rajput2021,stryker2019}. These methods have so far proven to be very specific to the target models and its symmetries. Thus the endeavor to find general, optimal mappings will be essential in the coming years for scaling up near term HEP simulations into the fault tolerant era. Comparing these methods to general fault tolerant encodings for universal QC will yield insight into the efficacy of building custom machines dedicated to HEP problems which implement custom encodings in hardware rather than reconfigurable devices capable of universal operations.

\subsubsection{Numerical Simulators}
Since numerical simulators will continue to play a vital role in quantum device verification and validation as well as in quantum algorithm analysis \cite{arute2019quantum,villalonga2020establishing}, the relevant research priorities are concerned with scaling numerical simulation techniques towards larger NISQ devices as well as improving their ability to faithfully reproduce the behavior of actual quantum hardware via more accurate noise models, in particular proper modeling of the multi-qubit cross-talk \cite{mccaskey2018validating}. In turn, devising more accurate noise models will require further research and development in scalable approaches for modeling open quantum system dynamics coupled with either a Markovian or non-Markovian bath. Further research related to simulations of multi-level qudit systems at their native Hamiltonian level will be necessary for assessing their computational power as compared to the qubit-based quantum devices. Pulse-level simulations, optimal quantum control, and quantum gate implementation optimization will form another direction of simulation heavy research and development efforts. More efficient numerical tensor-algebraic techniques coupled with advances in classical machine learning will be required for scalable characterization of larger NISQ devices. On the engineering side, all these newly devised numerical simulation techniques will have to be implemented in an efficient manner in order to fully exploit the computational power of Exascale HPC platforms which are becoming widely available worldwide \cite{britt2017quantum,doi.org/10.1049/qtc2.12024,nguyen2021scalable,humble2021quantum}.

\subsection{Experimental Hardware}
\subsubsection{Testbeds} 
It is in the interest of the HEP community to build specialized devices to address the above application needs. For example, multiple platforms offer analog-level operations to users (ie, pulse controls). In many scenarios, access to lower level, analog operations allows for higher accuracy in quantum simulation problems. Further, hybrid analog-digital simulations have shown increased accuracy in field theory problems for scattering calculations~\cite{arrazola_digital-analog_2016}.
Currently, HEP programs - especially those in the QuantISED program in the US - utilize testbeds around the world, including the DOE-supported testbeds consisting of trapped ion and superconducting devices, to solve prototypical problems of interest in the field. For example, the information scrambling problem~\cite{blok_quantum_2021} has been demonstrated on superconducting hardware \cite{Kreshchuk2021}, while variational methods for solving problems using the light-field encoding has been mapped onto trapped ion platforms \cite{Echevarria2021}. Uncovering the most useful aspects of the testbeds in these demonstrations will help us discover the most important ingredients for testbeds that support HEP goals. In particular, the "close to the metal", analog controls were a defining factor in these demonstrations.

In the NISQ era, fully analog systems will also be useful for implementing complex Hamiltonians for which digital, prefault-tolerant platforms do not support sufficient circuit depth. Hamiltonians for problems of interest can be mapped to trapped ion systems' motional mode degrees of freedom, or the spin-spin coupling present in neutral atom, quantum dot, or other platforms which support both analog spin-spin coupling and tunable transverse fields. In the future, when digital platforms move closer to fault tolerance, direct fermionic (HEP problem-specific) encodings should be supported in custom testbeds. These encodings ensure that the resources required in a given testbed required to represent a specific HEP problem are far fewer than that required for general purpose, universal quantum simulation. A custom, fault tolerant fermionic code may not be suitable to simulate Boson sampling, for example, but it would be ideal for simulating scattering in fermionic field theories.

\subsubsection{Analog Simulators}
While digital quantum computers scale to the extent needed for fault tolerant quantum computations, \textit{analog} quantum simulators are beginning to push state of the art, for example probing topological spin liquids on a programmable device at a scale not previously observed on neutral atom-based simulators \cite{semeghini_probing_2021,scholl_quantum_2021,bluvstein_controlling_2021}.
% 10.1126/science.abi8794, 10.1126/science.abg2530, https://www.nature.com/articles/s41586-021-03585-1.pdf}
Analog neutral atomic physics supplements photonic, superconducting, trapped-ion, and many more analog modalities which currently exist. There has also been recent research interest in exploring digital-analog computational modalities (in which hard operations are done in an analog fashion and digital operations play a complimentary supporting role, e.g. providing local basis transformations) and their application to HEP (and related condensed-matter) models as discussed in the previous section. 
Despite this promise, developing current analog methodologies to solve substantially more complex HEP domain problems is a research avenue in its own right. 

\subsubsection{HEP-dedicated Testbeds}

For HEP, an open and relevant question is whether quantum computers can efficiently simulate quantum field theories (QFT). QFT encompasses all fundamental interactions, possibly excluding gravity. The simulation of QFT has HEP applications related to event generators for quantum chromodynamics (QCD) simulations of nuclear matter. Outside HEP, it might find applications in the simulation of strongly coupled theories and the characterization of the computational complexity of quantum states.

Nonetheless, the simulation of quantum field theories is not a trivial task due to the sign problem. In physics, the sign problem is encountered in calculations of the properties of a quantum mechanical system with a large number of strongly interacting fermions or QFTs involving a non-zero density of strongly interacting fermions. Since the particles are strongly interacting, perturbation theory is inapplicable, and one is forced to use brute-force numerical methods. The sign problem is one of the major unsolved problems in the physics of many-particle systems, limiting progress in nuclear physics, preventing the ab initio calculation of properties of nuclear matter, and limiting our understanding of nuclei and neutron stars. In quantum field theory, it prevents the use of lattice QCD \cite{deforcrand2010simulating} to predict the phases and properties of quark matter \cite{Philipsen2012}.

An exciting and promising alternative to the simulation of QFT is based on the idea of adiabatic quantum computation (AQC) \cite{farhi2000quantum, Albash2018}. In general, AQC is as powerful as universal quantum computation when a non-stoquastic Hamiltonian is used \cite{Marvian2019OnTC}, which is the case of Hamiltonians dealing with the sign problem. One of the critical aspects of AQC is quantum annealing \cite{Kadowaki1998}, which is a metaheuristic algorithm to solve combinatorial optimization problems by changing the parameters adiabatically or even non-adiabatically. Measurements on currently available quantum annealers are done only on the standard computational basis \cite{Johnson2011}; thereby, the Hamiltonian is stoquastic. 

Thus, a quantum testbed that fully realizes or exploits the power of AQC is needed to show that quantum speedup can be achieved when non-stochastic Hamiltonians relevant to HEP are used. Some models experiencing first-order and second-order phase transitions using non-stoquastic Hamiltonians have been proposed and have demonstrated a quantum speedup \cite{Ikeda2020}. This natural and efficient way to simulate QFT on a quantum testbed will eliminate the need to reformulate HEP-relevant Hamiltonians (such as QCD) to fit a finite-dimensional Hilbert space amenable to circuit implementations.

Formulating the future of QCD studies as relying on a successful implementation of AQC provides an alternative to collider experiments, where large statistics are needed to understand rare processes. If we can simulate the QCD Hamiltonian we can study some observables in the context of information theory. For example, understanding the probabilistic parton distribution in terms of entangling entropy \cite{Kuvshinov:2017ncj}. In other words, in a collider experiment we have access to a probability distribution by collapsing the wave function to one of the states, by using a quantum-mechanical simulation we have access to the full density matrix.

\subsection{Infrastructures, Platforms and Data Networks}
\subsubsection{Software Development Frameworks, Benchmarks, and Optimisation}
The HEP community has developed over the years a broad range of algorithms and methods for different applications, from optimization to simulation and machine learning. Initial work on developing and assessing quantum equivalents of common algorithms have produced valuable insight \cite{Guan_2021}. However, the performance of NISQ systems is still limited and full advantage is expected only once reasonable fault tolerant devices will become available. To fully exploit the potential of quantum algorithms, software and hardware systems should be designed as part of co-development efforts across the HEP and quantum computing science and engineering communities. Experience in the classical domain shows that innovation in hardware technology, readily available to software and algorithm experts, can boost the development of new software stacks and algorithms that, in turn, produce critical feedback to hardware development in a virtuous cycle. Examples include hardware-aware compiler optimization, design of application-aware hardware architectures (i.e. systems dedicated to accelerating quantum simulations) and so on.
\par 
A fundamental component of measuring the progress and impact of quantum algorithms is the establishment of open benchmarking frameworks and platforms where different combinations of algorithms, software and hardware architecture can be tested, and the results published as a means of establishing baselines for further development. Such benchmarks can form the base for the development of common software tools and interfaces with an increasing level of abstraction and platform neutrality as typical today for classical computing resources. 
\par 
As an example, in 2021, CERN launched the ABAQUS project with the objective to build an open, easily accessible platform for researchers to become familiar with different quantum computing architectures and tools, produce or get access to benchmarks over different metrics, and build a community database of knowledge of the applicability of quantum algorithms to HEP workloads.
There are specific aspects that need to be investigated and studied. For example, there are easier simulations that are only polynomially hard, such as tensor networks with limited entanglement or limited number of non-Clifford gates that scale polynomially in the number of Clifford gates, but exponentially in the number of non-Clifford gates. It needs to be understood how to adapt and employ these simplifications. 
\subsubsection{Heterogeneous Computing Networks}
When dealing with the quantum computing paradigm, the role of infrastructure deserves a particular attention. Quantum computing should not over time require separate computing infrastructures, application platforms, or languages, but must be integrated into the existing cloud-based physical and software infrastructure of today and enable the deployment of hybrid, heterogeneous workloads where the quantum processing units (QPUs) are part of a broad range of accelerators. Presently classical resources can reliably store, manage and process huge quantities of data, while quantum devices are expected to efficiently explore high-dimensional spaces to extract insights and identify optimal answers.
\par
The development and deployment of such hybrid infrastructures is currently happening across a series of steps. As current NISQ resources are limited in availability and efficiency, experience is built by simulating quantum algorithms and building noise mitigation strategies on testbeds of classical resources of increasing scale. Since memory requirements double for every qubit added in the simulation and quickly hit an exponential wall, simulation of large scale systems can be performed by means of distributed and parallelised computation across HPC clusters of CPU/GPU. Novel architectural frameworks to integrate quantum capabilities APIs in cloud architectures are also emerging \cite{Grossi_2021}.
\par
At the same time, it is critical to start building paths to architecture optimization of hybrid resources, workload splitting and hybrid algorithms, scheduling, data flows, and circuits transpilation processes. Looking at hybrid models of variational classifiers, reinforcement-learning optimisation algorithms, quantum-classic GAN models for simulation, the infrastructure is evolving rapidly to make the classic and quantum interaction more efficient, defining containerised modules to be tested on distributed HPC environments and loaded on real quantum computers. This will need to be achieved by integrating HPC and quantum computing strategies as part of national and international initiatives.
\par 
One example of how such integration may be addressed is the Quantum Computing User Program (QCUP) at the Oak Ridge Leadership Computing Facility, a US Department of Energy user facility operated by Oak Ridge National Laboratory \cite{qcup}. QCUP offers access to a diversity of quantum computing systems following merit-based review of user proposal for demonstrate scientific applications. In addition, these users are capable of integrating their quantum computing workflows alongside conventional HPC systems. As quantum computing technology matures, this setting will offer a natural means by which to integrate quantum computing and conventional HPC systems for purposes of advanced computation.
\subsubsection{Quantum Data Networks}
A particular aspect of future Quantum/HPC infrastructures is the role of quantum data networks. Today quantum computing and quantum communication are usually handled as related by separate fields of applications of quantum technologies. However, quantum infrastructures, time and frequency distribution networks, and the development of quantum data technologies will play a fundamental and transformative role in physics experiments at different energy regimes (for example standard reference signals for anti-matter and low-energy experiments, such as ASACUSA and ALPHA, using high-precision laser spectroscopy \cite{Amsler:2748998}).
\par
Many national and international initiatives are being deployed to build future quantum networks and the HEP community must remain an active part of this development as it was at the beginning of the century in the development of distributed grid and cloud networks.
\subsubsection{International Collaborations and Co-Development}
The HEP community is distributed over the whole globe and the collaboration across institutes has always played a fundamental role in pushing the limits of science and technology. For example the development and operation of the Worldwide LHC Computing Grid (WLCG) was and still is instrumental for joint progress and employing all brain power wherever located in the world. The same approach needs to be taken in the development of future quantum infrastructures, Synergies between the Snowmass exercise and equivalent discussions in the European Particle Physics Strategy and other similar initiatives will be critical. Joint R\&D projects, testbeds, progressive development of common frameworks and tools across US Quantum Information Science institutes, the CERN Quantum Technology Initiative \cite{DiMeglio:2789149}, DESY QUANTUM, and many other national and international initiatives in EU and other countries will accelerate the development and adoption of quantum computing across the HEP community. In the same way, attention should be given to co-development not only across HEP, but also between HEP institutes, industry, and other scientific disciplines, such as Astrophysics, Cosmology, Earth Observation, Climate research, Condensed Matter, Chemistry, Biology, and many more. Often very similar challenges appear in different disciplines and can be solved with one solution.

%%%%%%%%%%%%%%%%%%%%%%%%%%%
\section*{Acknowledgements}
This work was supported by  supported by the U.S. Department of Energy, Office of Science, National Quantum Information Science Research Centers, Quantum Science Center, and the Advanced Scientific Computing Research program office Accelerated Research for Quantum Computing (ARQC) program. This research used resources of the Oak Ridge Leadership Computing Facility, which is a DOE Office of Science User Facility supported under Contract DE-AC05-00OR22725.

\bibliography{bibliography-ver-3.bib}

\end{document}